\documentclass[useAMS,usenatbib]{mn2e}
\usepackage{psfig}

\newcommand{\be}{\begin{equation}}
\newcommand{\en}{\end{equation}}
 
\newcommand{\lapp}{\mbox{\raisebox{-0.3em}{$\stackrel{\textstyle <}{\sim}$}}}
\newcommand{\gapp}{\mbox{\raisebox{-0.3em}{$\stackrel{\textstyle >}{\sim}$}}}

\def\zabs{$z_{\rm abs}$}
\def\zem{$z_{\rm em}$~}
\def\lya{Ly$\alpha$ }
\def\lyb{Ly$\beta$ }
\def\lyg{Ly$\gamma$ }

\def\h2{H$_2$}
\def\hi{H~{\sc i}~}
\def\ovi{O~{\sc vi}~}

\def\nv{N~{\sc v}~}

\def\oviab{O~{\sc vi}$\lambda\lambda$1031,1037~}
\def\ciii{C~{\sc iii}$\lambda$977~ }
\def\c2{C~{\sc ii}$\lambda$1036~ }

\def\civ{C~{\sc iv}~}
\def\civa{C~{\sc iv}$\lambda$1548~}

\def\civab{C~{\sc iv}$\lambda\lambda$1548,1550~}

\def\svi{S~{\sc vi}~}
\def\siv{S~{\sc iv}~}
\def\niv{N~{\sc iv}~}
\def\oiv{O~{\sc iv}~}
\def\neviii{Ne~{\sc viii}~}
\def\pv{P~{\sc v}~}
\def\sv{S~{\sc v}~}

\def\siiv{Si~{\sc iv}~}

\def\kms{km~s$^{-1}$}

\title[Outflowing material in the CSS quasar 3C48]{Outflowing material in the CSS quasar 3C48:
evidence of jet-cloud interaction?}
\author[Neeraj Gupta et al.]{Neeraj Gupta$^{1}$\thanks{E-mail:
neeraj@ncra.tifr.res.in (NG); anand@iucaa.ernet.in (RS); djs@ncra.tifr.res.in (DJS)},
R. Srianand$^{2}$ and D.J. Saikia$^{1}$ \\
$^{1}$National Centre for Radio Astrophysics, TIFR, Pune 411 007, India \\
$^{2}$Inter-University Centre for Astronomy and Astrophysics, Pune 411 007, India}
\begin{document}

\date{Accepted. Received; in original form }

\pagerange{0 -- 0} \pubyear{2005}

\maketitle

\label{firstpage}

\begin{abstract}
We report the detection of a \zabs=0.3654 associated absorption-line system 
in the UV spectrum of the CSS quasar 3C48.  The absorbing material is blue shifted 
with respect to the quasar emission-line redshift, \zem=0.3700, suggesting an outflow 
velocity of $\sim$1000 \kms.  
We detect absorption lines over a range of ionization states from Ly$\beta$, Ly$\gamma$, 
C~{\sc iv}, N~{\sc iv}, S~{\sc vi} to O~{\sc vi} and possibly O~{\sc iv} and Ne~{\sc viii}.  
The kinematical properties of the absorption-line system are similar to the 
blue-shifted emission line gas 
seen in [O{\sc iii}]~$\lambda$5007 (Chatzichristou, Vanderriest \& Jaffe 1999), 
which is believed to have interacted with the radio jet.  
We study the properties of the absorbing material using CLOUDY 
and find that photoionization models with Solar abundance 
ratios (with overall metallicity in the range 0.1$\le$Z/Z$_\odot$$\le$1.3) are enough to explain the 
observed column densities of all the species except \neviii, detection of which requires confirmation.  
Since the cooling and recombination time for the gas is $\sim$10$^5$ yr, the consistency with 
the photoionization models suggests that any possible interaction of absorbing material with the 
jet must have taken place before $\sim$10$^5$ yr. 
The abundance ratio of nitrogen to carbon is close to Solar values,
unlike in the case of most quasars, especially at high-redshifts, which have super-Solar values. 
We observed 3C48 with the Giant Metrewave Radio Telescope (GMRT) to search for 
redshifted 21cm \hi absorption. However, we did not detect any significant feature in our spectra and 
estimate the 3$\sigma$ upper limit to the optical depth to be in the range 0.001 to 0.003.  
However, due to the diffuse nature of the radio source, optical depths as 
high as 0.1 towards individual knots or compact components cannot be ruled out.
\end{abstract}


\section{Introduction}
\begin{table*}
\caption{Observational log of UV data}
\begin{center}
\begin{tabular}{lcccccc}
\hline
\hline
Program &Instrument&      PI        & Exposure & Wavelength & Resolving
&Signal-to- \\
ID      &used      &                & time     & coverage   & power &noise
\\
        &          &                &  (s)     &  (\AA)     &  (R) &(per
spectral resolution)\\
\hline
9280    &HST STIS  &A. Siemiginowska&8,113     &2049$-$2139   & 11,000    & 7\\
8299    &HST STIS  &W. Jaffe        &5,280     &8540$-$9107   & 8,500    & 20\\
6707    &HST GHRS  &D. Bowen        &21,760    &1287$-$1575   & 2,000     &15$-$25    \\
Z002    &FUSE      &P. Wannier      &8,347     &985$-$1084    & 20,000    & 7$^a$    \\
KQ003   &IUE       &O'Brien         &24,960    &1150$-$1979   & 250       & 4  \\
\hline
\end{tabular}
\end{center}
$^a$ The FUSE spectrum has been smoothed over 15 pixels. 
\label{uvtab}
\end{table*}
Compact steep-spectrum sources (CSSs) are defined to be 
radio sources with projected linear size $\lapp$20 kpc 
(H$_o$=71 km s$^{-1}$ Mpc$^{-1}$, $\Omega_m$=0.27, $\Omega_\Lambda$=0.73,
Spergel et al. 2003) and having a steep high-frequency radio spectrum
($\alpha\gapp$0.5, where F($\nu)\propto\nu^{-\alpha}$).  High-resolution
radio images of CSSs show that there is a wide variety of structures, ranging
from double-lobed and triple sources to those which are very complex and
distorted (e.g. Phillips \& Mutel 1982; Spencer et al. 1989; Fanti et al. 1990;
Conway et al. 1994; Wilkinson et al. 1994; Sanghera et al. 1995; Dallacasa et al. 1995;
Readhead et al. 1996a; Taylor et al. 1996; Snellen et al. 2000a; Stanghellini et al. 2001;
Orienti et al. 2004).  Although there 
is a consensus of opinion that most CSSs are young radio sources (Fanti
et al. 1995; Readhead et al. 1996b; O'Dea 1998 for a review; Snellen et al. 2000b),
a small fraction could be confined to small dimensions due to either
a dense medium or interactions of the jet with dense clouds. 
Radio studies have shown that CSSs tend to be more 
asymmetric than the larger sources, possibly due to interaction with 
an asymmetric distribution
of gas in the central regions (Saikia et al. 1995, 2001; Saikia \& Gupta 2003). These authors
speculated that this might be due to collisions with clouds of gas, some of
which fuel the nuclear activity. Arshakian \& Longair (2000) also concluded
that intrinsic/environmental asymmetries are more important for sources of
small physical sizes. Radio polarization studies often indicate large Faraday depths
(e.g. Mantovani et al. 1994; Fanti et al. 2004), and also sometimes show evidence 
of asymmetrically-located clouds as a huge differential rotation measure between the oppositely
directed lobes as in the CSS quasar 3C147 (Junor et al. 1999), or 
high rotation measure where the jet bends sharply suggesting collision of the 
jet with a dense cloud as in 3C43 (Cotton et al. 2003). 
Optical studies of CSSs have also often shown evidence of interaction of the 
radio jets with the external medium (e.g. Gelderman \& Whittle 1994; de Vries et al. 1999; 
Axon et al. 2000; O'Dea et al. 2002).
The physical conditions under which a jet may be bent or distorted by such
clouds, as well as the effects of an asymmetric gas distribution on opposite
sides of the nucleus have been explored via analytical calculations as well as 
numerical simulations (Carvalho 1998; Higgins, O'Brien \& Dunlop 1999; 
Wang, Wiita \& Hooda  2000; Jeyakumar et al. 2005).

\begin{figure*}
\centerline{\vbox{
\psfig{figure=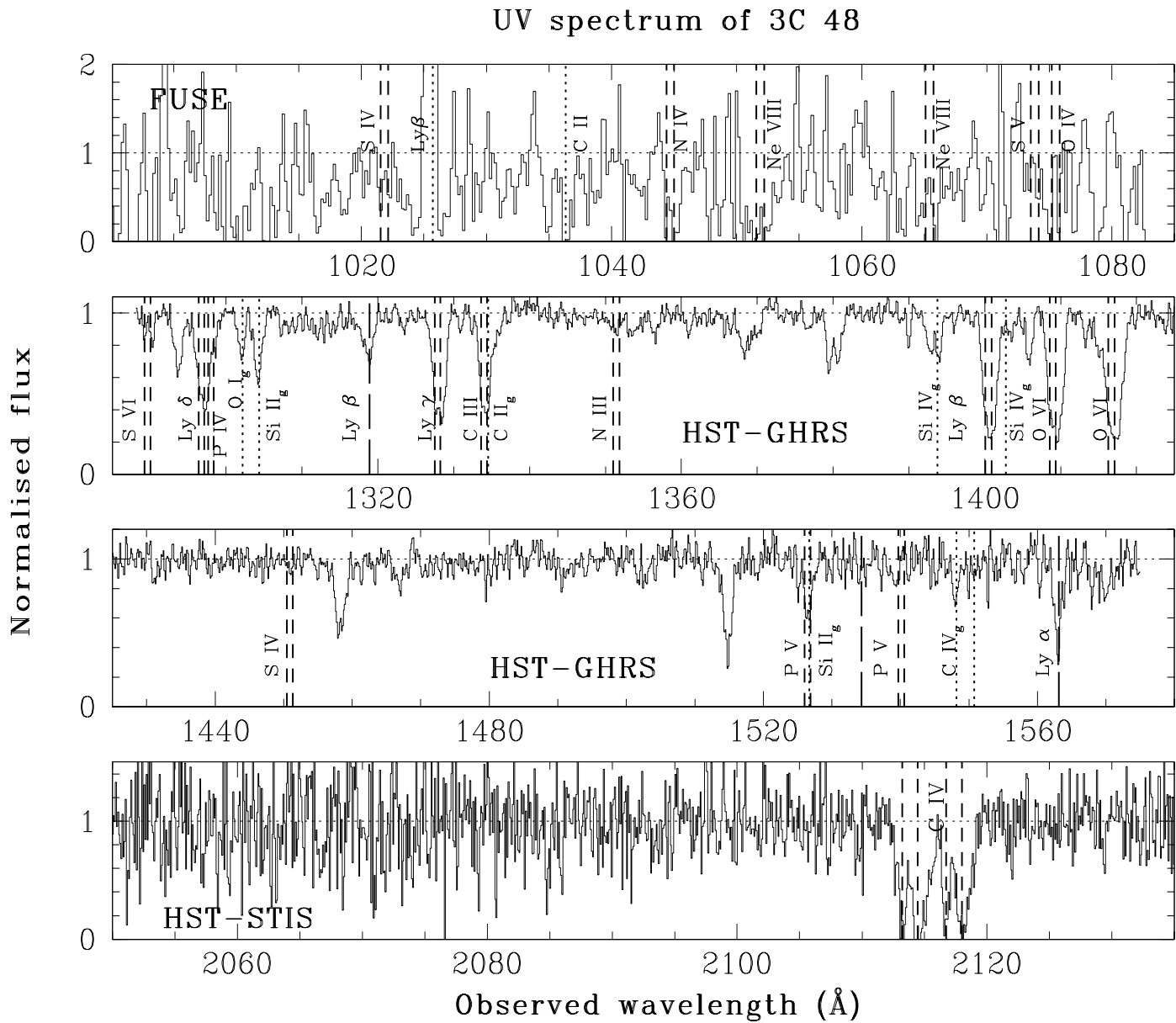,height=16.5cm,width=20.0cm,angle=-90}
}}
\caption[]{UV spectrum of the \zem = 0.3700 quasar 3C48.  Different absorption 
lines are labelled.  Vertical dashed lines represent the positions of the
absorption components associated with the \zabs=0.3649 and 0.3658 systems.  
Short-dashed lines with the label `g' correspond to Galactic absorption.  
Long dashed lines mark a possible intervening system, \zabs=0.286.  
A number of unidentified absorption features (left unmarked) are also 
present in the spectrum. It is most likely that these are intervening \lya 
absorption lines.
}
\label{fspec}
\end{figure*}
An interesting way of investigating the interaction of the jet with the 
external clouds/medium could be via the absorption lines arising from the clouds
accelerated by the jet. Baker et al. (2002) have studied the absorption spectra
in a sample of quasars selected from the Molonglo Reference Catalogue, and find 
a slight excess of blue-shifted \civ absorption lines in CSSs compared with the
larger objects. In a sample of high-redshift radio galaxies, van Ojik et al. (1997)
reported a high incidence of \hi absorption (9 out of 10) in the smaller objects
($<$50 kpc) compared with the larger ones (2 out of 8).  Their study also showed
an excess of blue-shifted \lya absorption lines, and showed strong evidence of
interaction of the radio source with the external environment.
Although blue-shifted absorption lines could
arise due to either halo gas or circumnuclear gas kinematically affected by 
nuclear winds and/or radiation pressure, such studies suggest 
that jet-cloud interaction may also play a significant role in these objects. 

To explore this theme, we have studied the well-known CSS quasar 3C48, which has a
highly complex and distorted radio structure, speculated to be due to disruption
of the jet by interaction with a cloud (e.g. Wilkinson et al. 1991; Worrall et al. 
2004; Feng et al. 2005).
We explore evidence of blue-shifted absorption line gas at radio and ultraviolet
(UV) wavelengths using the Giant Metrewave Radio Telescope (GMRT) and archival, 
but previously unpublished, 
data from the Hubble Space Telescope (HST) 
Space Telescope Imaging Spectrograph (STIS) and the Goddard High Resolution 
Spectrograph (GHRS), the Far Ultraviolet Spectroscopic Explorer (FUSE) and the 
International Ultraviolet Explorer (IUE). We report the detection of blue-shifted
absorption line gas, examine the physical properties of the absorber and discuss
the possibility that this might be accelerated by interaction with the radio jet.

\section{3C48}
3C48, the second quasar to be identified (Mathews \& Sandage 1963), is an enigmatic 
CSS source. It has unusually strong IR emission (Neugebauer, Soifer \& Miley 1985), 
comparable with most 
luminous IR galaxies. The detection of CO emission (Scoville et al. 1993; Wink, 
Guilloteau \& Wilson 1997) 
at z=0.3695 implies the presence of about twice as much molecular gas as in ULIRGs.  
At radio wavelengths, the VLBI images resolve this highly compact radio source into a 
one-sided core-jet structure confined well within the optical host galaxy. Briggs (1995) 
and Feng et al. (2005) have reported the detection of a weak counter-jet on subarcsec scales.  
Wilkinson et al. (1991) interpret the complex structure of 3C48 as due to collision of 
the jet with a dense clump of gas in the host galaxy's interstellar medium.  The radio 
jet is also aligned with the excess optical continuum, and [O{\sc ii}]~$\lambda$3727 and 
[O{\sc iii}]~$\lambda\lambda$4959,5007 emission-line nebulosity seen close to the quasar core, 
and in particular the peak of continuum light at about 1 arcsec from the nucleus.  
This has been variously interpreted as the nucleus of a galaxy, merging with the host of 
3C48 (Stockton \& Ridgway 1991) or a star-forming 
region triggered by the radio jet (Chatzichristou et al. 1999).

The redshift of 3C48 as measured by several authors (Greenstein \& Mathews 1963; Wampler et al. 1975; 
Thuan, Oke \& Bergeron 1979) differ by about $\sim$500 \kms.
In particular, Boroson \& Oke (1982) noted that redshift of 3C48 as determined from the forbidden 
emission lines turns out to be slightly less than the value of 0.3700$\pm$0.0002 determined 
from the permitted emission
lines.  The discrepancy arose largely due to the lower spectral resolution data.  In the higher 
resolution data of Chatzichristou et al. (1999), the forbidden lines are resolved into two 
components, one of which shows velocities similar to the permitted lines.  The value of 0.3700 
is also in agreement with the emission lines and stellar absorption lines seen from the fuzz around the 
source.  The MgII emission line (Tytler et al. 1987)  and CO emission (Scoville et al. 1993) have 
also been seen at redshifted frequencies corresponding to $\sim$0.370.  Therefore, throughout this 
paper we use \zem=0.3700 to represent the quasar redshift.

\section{Observations and data analyses}
We first describe the analyses of the archival but previously unpublished UV data from
HST STIS and GHRS, FUSE and IUE. These data have been obtained from
the Multimission Archive at Space Telescope (MAST). We then describe the
radio observations with the GMRT to detect redshifted \hi in absorption towards
3C48.

%
\subsection{UV data}
%
\begin{figure}
\centerline{\vbox{
\psfig{figure=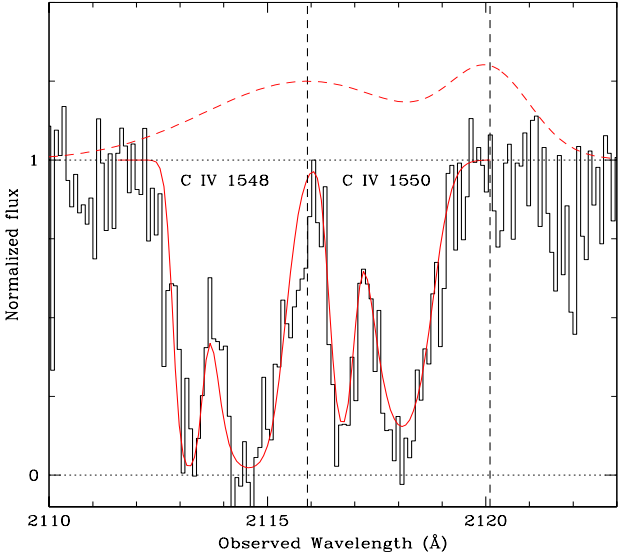,height=8.0cm,width=9.0cm,angle=0}
}}
\caption[]{Profile of \civab absorption lines detected in the STIS
spectrum of 3C48 (histogram) overplotted with the best-fitted Voigt profiles
(smooth curves). 
Two well-detached absorption components are detected clearly at 
\zabs= 0.3649 and 0.3658. Vertical long-dashed lines mark
the expected location of \civa emission based on the two velocity 
components detected in [O{\sc iii}]$\lambda\lambda$4959,5007 and 
[O{\sc ii}]$\lambda$3727 emission lines 
(Chatzichristou et al. 1999). The figure also shows 
the expected velocity range of emitting gas (Gaussians with
long-dashed lines) reconstructed using the parameters of Chatzichristou et al. (1999).
The gas producing the \civ absorption is well
detached from the nuclear line-emitting gas and has
a velocity consistent with the blue component of the narrow line-emitting gas.
}
\label{emabs}
\end{figure}
The data from HST STIS and GHRS, FUSE and IUE instruments were reduced using
the standard pipeline tasks {\tt calstis, calhrs, calfuse and caliue} respectively.  
The pipeline processing keeps track of
errors associated with each pixel at every stage of the reduction.
The spectra corresponding to various exposures for each instrument
along with the error spectra were then combined separately.  The resulting
signal-to-noise ratio applicable over most of the wavelength range
and some of the basic details of the data sets used are summarised in
Table~\ref{uvtab}.  The spectra for each of the data sets 
were then continuum fitted using absorption free regions to get the final 
spectrum shown in Fig.~\ref{fspec}.

The pipeline tasks also process the line lamp exposures to determine the
zero point offset of wavelength and spatial scales in the image.
We also made use of the Galactic absorption lines detected in each of the
spectra to determine the zero point offsets.  In particular, for the GHRS
spectrum we find that  it is required to shift the spectrum by approximately
+0.8 \AA.  This is also consistent with the rms of approximately one diode
(roughly 0.6 \AA) in the default wavelength scale of GHRS pipeline calibration. 
The one diode rms in the wavelength scale arises due to the errors in the 
carrousel positioning and uncertainties caused due to the changing temperatures 
within the spectrograph and geomagnetically induced image motions (see {\it Instrument Handbook 
for the GHRS} for details). 
The spectrum corresponding to three GHRS exposures were combined after applying 
this offset.

\subsection{Radio observations with GMRT}
We observed 3C48 with GMRT to investigate redshifted 21cm absorption with a bandwidth
of 4 MHz in the 1060 MHz subband. The  FX correlator system at GMRT splits
the baseband bandwidth into 128 spectral channels, yielding a spectral 
resolution of 8.75 \kms for the chosen bandwidth. We observed 3C48 on a
number of occasions with the redshifted 21cm frequencies 
corresponding to \zabs = 0.3654, the redshift at which we detect outflowing
gas (Section 4) and 0.3695 and 0.3700, the emission-line redshifts from CO and 
optical observations respectively. This object was observed as part of a larger survey of
\hi absorption towards CSSs, and the observational details will be described in more
detail elsewhere. The radio data were reduced in the standard way using the
Astronomical Image Processing System.
%
\section{Observational results}
\subsection{UV spectra}

We detect blue-shifted absorption lines with \zabs$\simeq$\zem in the UV spectra of
3C48. Fig.~\ref{fspec} shows the normalised UV spectrum obtained with
HST and FUSE. Expected positions of absorption lines from our Galaxy, 
the associated system at \zabs = 0.3654 and an  intervening 
\lya system at \zabs = 0.286 are marked with short-dashed, dashed and long-dashed
vertical lines respectively. In the case of the \zabs = 0.3654 system we have marked the 
locations of the two velocity components identified from the \civab absorption lines
seen in the STIS spectrum. 
It is clear from the figure that Ly$\beta$, Ly$\gamma$, Ly$\delta$, 
C~{\sc iv}, \ovi and S~{\sc vi} absorption lines from the \zabs = 0.3654 
system are present in the HST-STIS and GHRS spectra. The \ciii 
absorption line is possibly blended with C~{\sc ii}$\lambda1334$ absorption 
from our Galaxy. We find Ly$\delta$ to be slightly stronger than 
the Ly$\gamma$ line, which is possibly because of blending with 
the P~{\sc iv} absorption line.

\par
Absorption lines of P~{\sc v} doublets and S~{\sc iv} are 
clearly absent. The available spectra do not cover the wavelength 
range where Ly$\alpha$, \nv and \siiv absorption lines are expected.
The FUSE spectrum is noisier and we have examined the different 
exposures independently to verify the reliability of absorption 
lines. We have detected \niv and possibly \oiv and \neviii.  To our 
knowledge, \neviii absorption has been reported for 5 other systems, 
namely 3C288.1, UM 675, HS 1700+6416, J2233$-$606 and SBS 1542+531
(Petitjean, Riediger \& Rauch 1996; Petitjean \& Srianand 1999; 
Hamann et al. 1997; Telfer et al. 1998; 
Hamann, Netzer \& Shields 2000). Based on the presence of absorption due to 
very highly ionized species and \zabs $\simeq$ \zem 
it is most likely that the gas responsible for the absorption 
is located close to the nuclear region of 3C48.

\par
The \civab absorption doublet detected in the STIS data is clearly 
resolved into two components at \zabs=0.3649 and 0.3658 
(Fig.~\ref{emabs}). Voigt profile decomposition gives,   
N(C~{\sc iv})=3.44$\pm$0.55$\times10^{14}$ cm$^{-2}$ , b =39.8$\pm$4.3 \kms  
and N(C~{\sc iv})=7.91$\pm$0.74$\times10^{14}$ cm$^{-2}$,   b =92.6$\pm$5.9 \kms 
respectively for these components. Within the errors of the STIS 
spectrum, it cannot be ruled out that the \civab absorption lines are not saturated.
Therefore, we take the column density estimates from the Voigt profile fits as a
lower limit. In Fig.~\ref{emabs}, alongwith the Voigt-profile fits to \civ absorption 
line, we show the location of \civa emission as expected from the two components 
detected in the [O{\sc iii}]~$\lambda\lambda$4959,5007 and 
[O{\sc ii}]~$\lambda$3727 emission lines (Chatzichristou et al. 1999).  
From the two-component profile decomposition of the emission-line spectrum 
integrated over the central 1.9 arcsec$^2$, Chatzichristou et al. (1999) 
find that one of the components (referred to as the red component) have 
redshift similar to the systemic redshift of the quasar while the other 
component (referred to as the blue component) is blue shifted by 
$\sim$580$\pm$15~\kms.  The systemic red component is much narrower 
(FWHM $\sim$400~\kms) and shows a systematic trend in the velocity structure 
suggesting association with the rotational velocity field of the galaxy.  
The blue component on the other hand has much wider (FWHM $\sim$1010~\kms) 
line profile and shows no clear trend in the velocity field structure.
From Fig.~\ref{emabs} it is clear that
(i) the absorbing gas completely covers the background source
and (ii) the velocity range spanned by the absorbing gas is 
consistent with that of the outflowing gas responsible for the 
blue-shifted [O{\sc iii}]~$\lambda\lambda$4959,5007 and [O{\sc ii}]~$\lambda$3727 
emission detected by Chatzichristou et al. (1999).

\par
In the lower-resolution GHRS 
data, the two absorption components detected in the STIS spectrum are unresolved.
However the widths of the detected absorption lines are consistent with
the presence of two components that are detected in the \civ profile.  
To keep the analysis simple, we do not try to get the column
densities of individual components by fitting two components to the
low dispersion data. Instead we estimate the total integrated 
column density in both the components.

\par
For the absorption lines detected in the GHRS and FUSE data sets, 
we estimate column density at each velocity pixel using the relation 
\begin{equation}
 N(v)=3.768\times 10^{+14}~\tau(v)/f\lambda~{\rm cm^{-2}~km^{-1}s}
\label{eqcol}
\end{equation}
where, $\tau$, $f$ and $\lambda$ are the optical depth, oscillator
strength and rest wavelength respectively. The latter two parameters
have been taken from Verner et al. (1994).    The total column
density is obtained by integrating $N(v)$ over the velocity range
over which absorption is detected (see Table~\ref{tcol}).   
The typical 3$\sigma$ errors are 25$\%$ of the estimates.  

Although absorption is a direct indicator of the optical depth, in 
the case of partial coverage of the background source or when optical depth 
varies much more rapidly as a function of frequency than the 
spectral resolution, the depth of the absorption line provides 
only an under-estimate of the true column density.  In case of poor 
spectral resolution, it is possible that the absorption in many cases may 
be actually optically thick.  For example, for \oviab doublet detected in 
the GHRS data, the column density estimates for the  members of the doublet 
just scale by their respective oscillator strengths.  This suggests that 
the \ovi absorption lines are actually saturated.  Therefore, we take the 
estimated value based on O~{\sc vi}$\lambda1036$ as a lower limit.   

Similar to \ovi, it is possible that neutral hydrogen column density 
estimated from \lyb and \lyg are at best the lower limits only.  
In the IUE spectrum, there is no apparent drop in the flux density at $\sim$1250 \AA~ 
corresponding to the Lyman edge absorption at the redshift of the 
quasar (\zem=0.3700) and the absorber (\zabs=0.3654).  
By taking 20 \AA~ on the blue side and 50 \AA~  on the red side of the 
expected Lyman edge, we find N(H~{\sc i})$\le$3.06$\times10^{16}$ 
cm$^{-2}$.  This is consistent with the N(H~{\sc i}) values listed
in Table~\ref{tcol} based on \lyb and \lyg. Thus the actual
value of N(H~{\sc i}) is estimated to be between $\sim5\times10^{15}$ cm$^{-2}$
and 3$\times10^{16}$ cm$^{-2}$.

We evaluate the upper limits for column densities of the species that
are not detected using the \lyg absorption profile as the template
(cf. Gupta et al. 2003). The scaling factor 
$k$~=~[$Nf\lambda$]$_{X^+}$/[$Nf\lambda$]$_{template}$ between the two 
optical depths $\tau$$_{template}$ and $\tau$$_{X^i}$ for species $X^i$ 
is then obtained by minimizing
\begin{equation}
\alpha = \Sigma(\tau_{\rm X^i} - k*\tau_{template})^{2}
\end{equation}
The upper limits calculated in this manner are tabulated in Table~\ref{tcol}.  
%
\begin{table}
\caption{Column density estimates for the various species.
}
\begin{center}
\begin{tabular}{lcc}
\hline
\hline
 Species & N & Instrument \\
         & (cm$^{-2}$) &     \\
\hline
\lyb     & 2.57$\times10^{15}$       &   HST-GHRS \\
\lyg     & 5.50$\times10^{15}$       &      "     \\
Lyman-   & $\le3.06\times10^{16}$    &   IUE      \\
Limit \\
\civ     & $\ge$1.14$\times10^{15}$       &   HST-STIS \\
\svi     & 8.62$\times10^{13}$       &   HST-GHRS \\
\ovi     & $\ge1.58\times10^{15}$    &      "     \\
\siv     & $\le9.32\times10^{13}$    &      "     \\
\pv      & $\le2.91\times10^{13}$    &      "     \\
N~{\sc iii}    & $\le1.30\times10^{14}$    &      "     \\
\niv     & 5.15$\times10^{14}$       &   FUSE     \\
\neviii  & 5.45$\times10^{15}$       &      "     \\
\sv      & $\le4.01\times10^{13}$       &      "     \\
\oiv     & 3.86$\times10^{15}$       &      "     \\
  
\hline
\end{tabular}
\end{center}
\label{tcol}
\end{table}

\subsection{21~cm absorption}
The column density of \hi derived towards the optical point source is very
low and the gas will not produce any detectable 21 cm absorption. A tentative 
detection of \hi absorption (with a FWHM of 100 \kms) towards 3C48 was reported
by Pihlstr\"{o}m (2001).  High-resolution VLBI observations show that 
most of the radio emission is extended (Wilkinson et al. 1991; 
Worrall et al. 2004). Only 10\% of the flux density at 1.53 GHz is in the collimated 
core-jet structure with only 1\% of the flux density being associated with the radio core that 
is coincident with the optical point source.  Thus the tentative detection 
of 21 cm absorption by Pihlstr\"{o}m (2001), if true, has to be towards the 
extended radio emission.

\par
\begin{figure}
\centerline{\vbox{
\psfig{figure=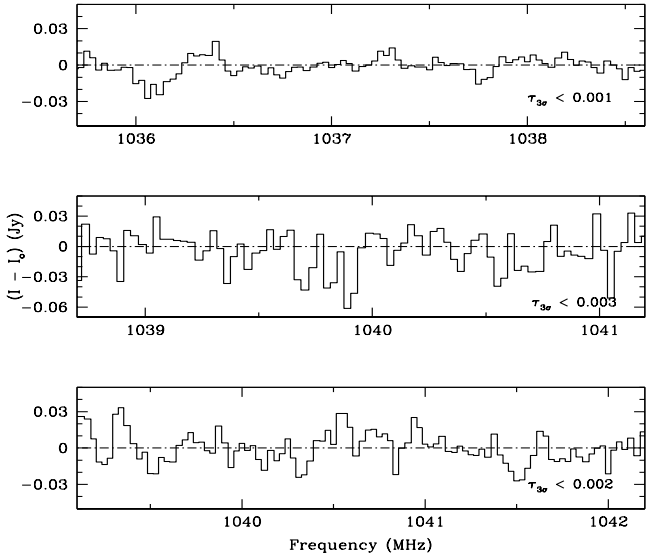,height=8.0cm,width=9.0cm,angle=0}
}}
\caption[]{GMRT spectra covering the redshifted 21cm frequency corresponding 
to \zem=0.3700 from the optical emission lines, \zem=0.3695 from the CO emission spectrum 
(upper panel), 
and  \zabs=0.3654 corresponding to the blue-shifted absorbing material reported here
(middle and bottom panels).}
\label{21cm}
\end{figure}

Our deep GMRT observations fail to reveal any detectable
21 cm absorption at the systemic redshift of 3C48 as well as the
gas responsible for UV absorption lines (Fig.~\ref{21cm}).
At the resolution of the GMRT observations, 
3C48 is unresolved. Using the flux density of 20.3 Jy 
estimated from our observations, we obtain the 3$\sigma$ upper limit 
on optical depth to be in the range of 0.001 to 0.003. We obtain an upper limit
on average N(H~{\sc i}) across the radio source using
\begin{equation}
N(H~{\sc I})=1.82\times10^{18}T_{s}\tau_{peak}\Delta V {\rm cm^{-2}}.
\label{eq21cm}
\end{equation}
Here $\tau_{peak}$ is the peak optical depth, $\Delta V$ the line FWHM in 
\kms, $T_{s}$ the spin temperature in K. Using the measured 3$\sigma$ 
upper limit on $\tau$ (instead of $\tau_{peak}$) 
we get N(H~{\sc i})$\lapp$1.8$\times10^{19}$cm$^{-2}$ for $\tau_{3\sigma}$ = 0.001 and 
assuming $T_{s}$=100 K and 
$\Delta V$=100 \kms. This is consistent with low N(H~{\sc i}) measured
towards the optical point source.
However, it should be noted that due to the extended nature of the
radio source, low-resolution observations will fail to detect the
presence of high N(H~{\sc i}) gas in front of say 
knot B (at about 0.3 kpc from the compact core A: see Worrall et al.
2004), near which the jet gets disrupted. Knot B    
has a 1.53 GHz flux density of 250 mJy that is $\sim2\%$ of the total
flux density.  This then implies that log $N$(H~{\sc i})~(cm$^{-2}$) $<$ 21.30 
along our line of sight towards knot B. 
Thus our GMRT observations cannot rule out the presence of high column density
\hi towards individual knots or compact features.
%

\section{Discussion}
The UV spectra show clear evidence of blue-shifted gas with 
\zabs=0.3654.  Compared with \zem=0.3700 this corresponds to a 
modest outflow velocity of $\sim$ 1000 \kms.  
\begin{figure}
\centerline{\vbox{
\psfig{figure=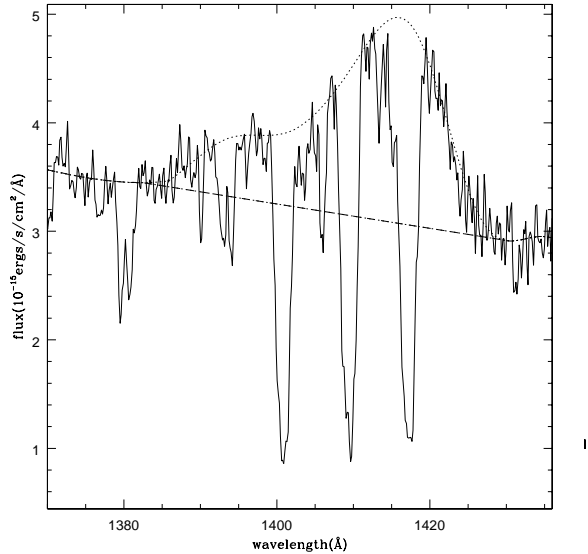,height=8.0cm,width=8.0cm,angle=0}
}}
\caption[]{Portion of the HST GHRS spectrum showing \lyb and \oviab absorption lines 
detected on top of the \ovi emission line.  
The low-order polynomial used to fit the continuum to the spectrum and model the 
emission line is shown as the dotted line. 
Dashed line represents the fit to the underlying continuum emission.} 
\label{covel}
\end{figure}
We can further constrain the location of the outflowing material by 
comparing the absorption features with the emission lines.
From Fig.~\ref{emabs}, we can conclude that the members of the \civab 
doublet have optical depths consistent with a full coverage of the background 
source.  As \civ, Ly$\beta$ and O~{\sc vi} 
absorption lines are detected on top of the emission lines
(see Fig.~\ref{covel}) complete coverage means that the gas is
outside the broad-line region (BLR) and has a projected size more than that of
the BLR (see Srianand \& Shankaranarayanan 2000).
We estimate the radius of BLR (R$_{BLR}$) using the empirical relation 
\begin{equation}
{\rm R}_{BLR}=27.4\bigg{(}\frac{\lambda L_{5100}}{10^{44}{\rm ergs~s^{-1}}}\bigg{)}^{0.68} {\rm lt~days}
\label{rblr}
\end{equation}
obtained by Corbett et al. (2003) using the reverberation mapping technique.
The quasar rest frame luminosity at 5100 \AA ~(i.e. L$_{5100\AA}$) is 
calculated using the observed flux density of 
F(8780 \AA)=8.03$\times$10$^{-16}$ ergs s$^{-1}$ \AA$^{-1}$ cm$^{-2}$, estimated from the STIS data set 
(05AB1030, program ID 8299; see Table~\ref{uvtab}) and a spectral index of 1.0. 
Substituting the numerical value of 
$\lambda$L$_{5100\AA}$ = 3.28$\times$10$^{45}$ ergs s$^{-1}$ 
in equation~(\ref{rblr}), the radius of BLR is estimated to be 0.25 pc.
Therefore we can conclude that size of the absorbing cloud should be larger 
than 0.5 pc and at least 0.25 pc away from the central engine. 

In the following subsections we discuss the physical properties of the 
outflowing absorbing gas and explore the possibility that these might be 
affected by the interaction with the radio jet.
%
\subsection{Physical conditions of the absorber}
\begin{figure}
\centerline{\vbox{
\psfig{figure=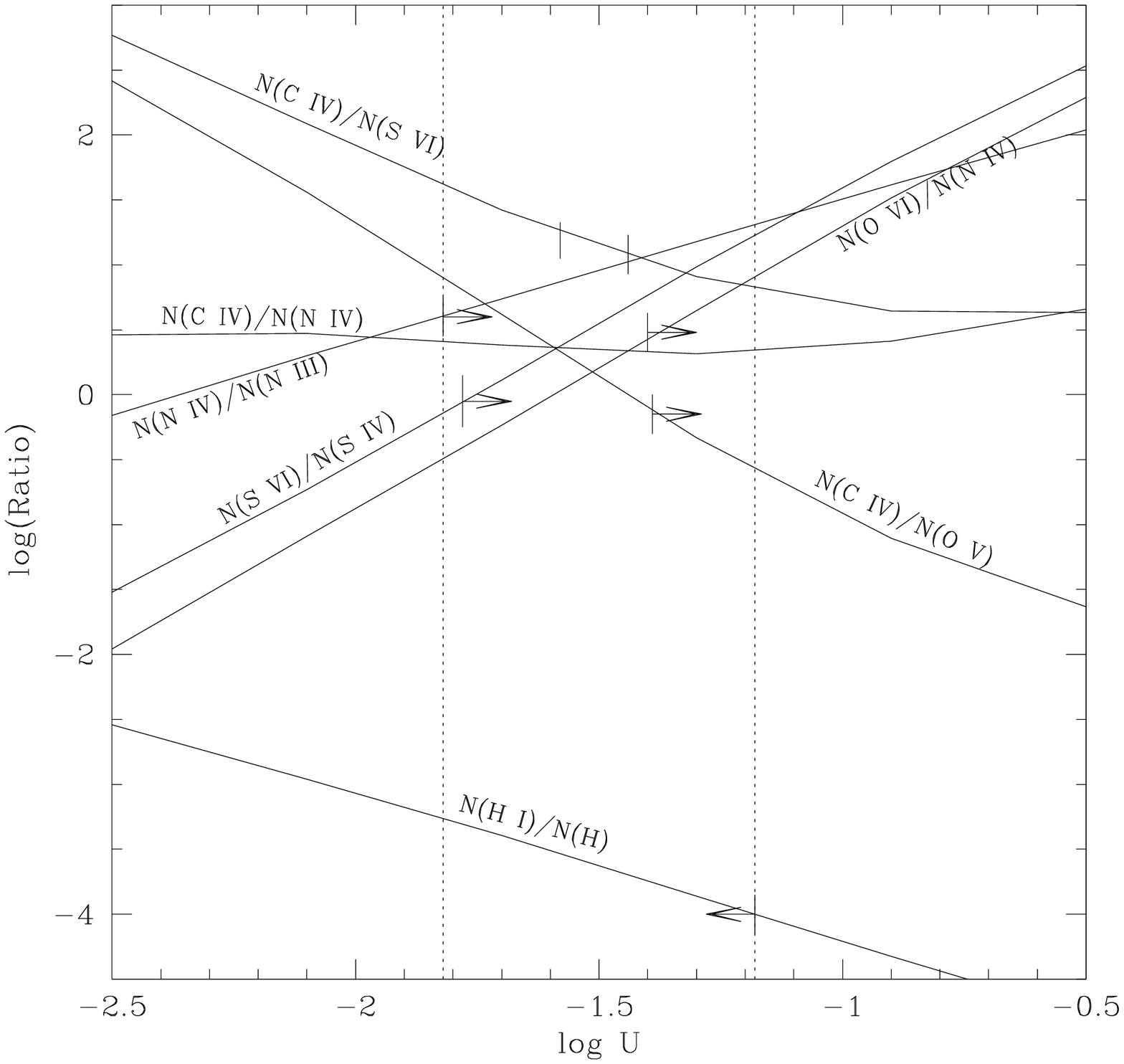,height=8.0cm,width=8.0cm,angle=0}
}}
\centerline{\vbox{
\psfig{figure=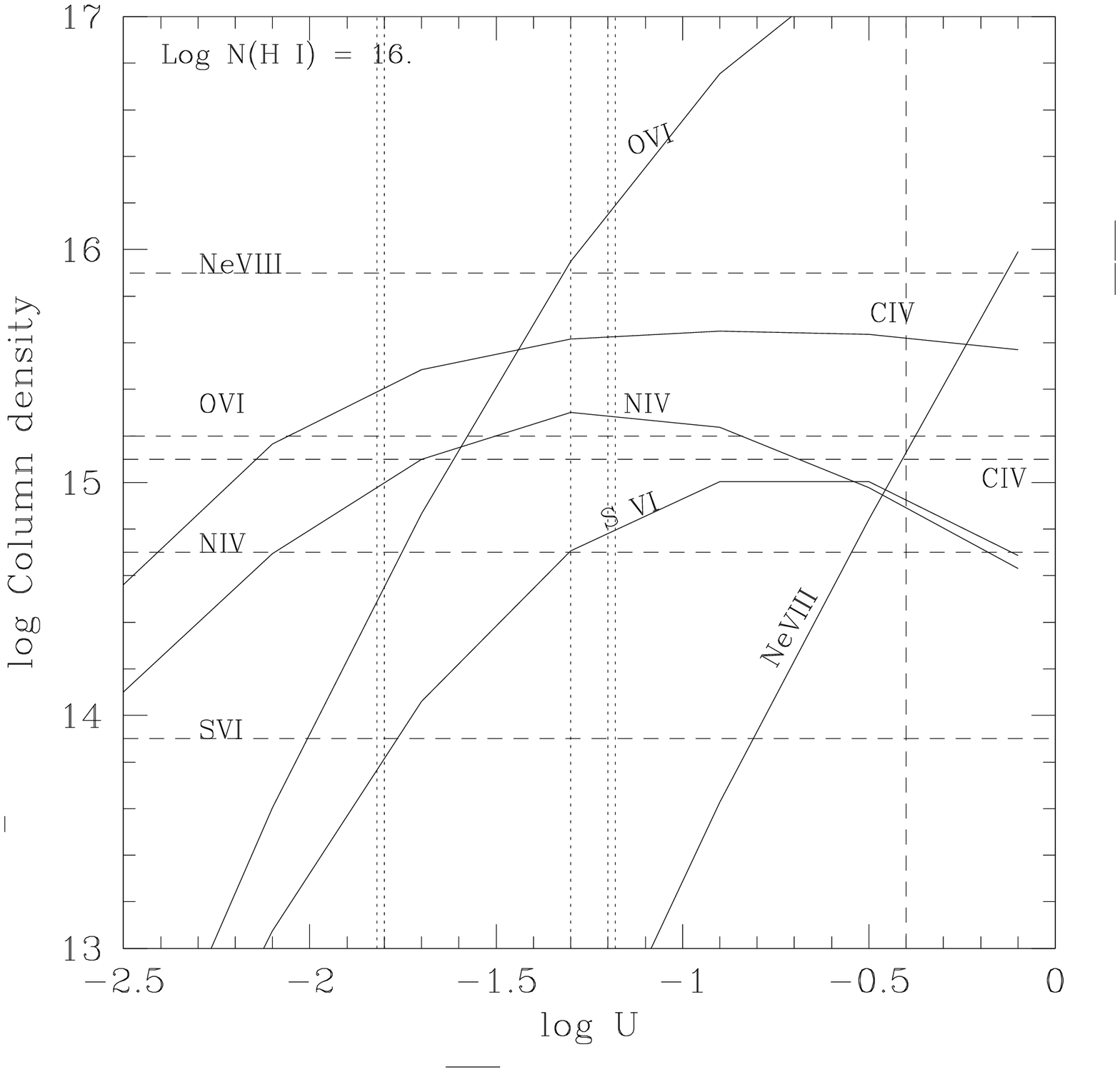,height=8.0cm,width=8.0cm,angle=0}
}}
\caption[]{Results of model calculations with N(H~{\sc i}) = 10$^{16}$ cm$^{-2}$,
QSO spectrum given by MF and Z = Z$_\odot$. Top panel shows different ion ratios
predicted in the calculations as a function of ionization parameter. The arrows
indicate the limits on U from the observed values. The vertical dashed lines
mark the possible range in U that will be consistent with the observations.
Lower panel shows the predicted column density as a function of U for a few ions.
The horizontal lines mark the measured values or upper limits to the column densities
(see Table.~\ref{tcol}). Apart from N(Ne~{\sc viii}) all the observed 
column densities and the limits are consistent with $-1.82\le {\rm log~U}\le-1.18$ and
Z = 0.5 Z$_\odot$ for log N(H~{\sc i}) = 16 cm$^{-2}$.  
}
\label{modfig}
\end{figure}

The observed column densities of different species can be used to obtain 
physical conditions of the absorbing gas.  As \zabs$\approx$\zem for the outflowing 
material, it is most likely that the ionization state of the gas will be influenced by the 
radiation field of the quasar.  In addition, jet-cloud interaction can also play an important 
role in determining the observed kinematics and ionization state of the gas in radio sources.  
First we consider the case of the absorbing gas in photoionization equilibrium 
with the UV radiation from 3C48.  For this purpose, we run grids of 
photoionization models using CLOUDY (Ferland 1996; Ferland et al. 1998) considering the 
gas to be ionised by a Mathews \& Ferland (1987; hereinafter referred to as 
the MF) spectrum and having Solar abundances.  The neutral hydrogen column 
density used is log $N$(H~{\sc i}) = 16.0 cm$^{-2}$.  We present the results of our 
calculations in Fig.~\ref{modfig} as a function of the dimensionless quantity, 
the ionization parameter U, which is defined as 
\begin{equation}
{\rm U} = \frac{Q(H^o)}{4\pi r^2cn_H}
~~~~~~ {\rm where}
~~~~~~ Q(H^o) = \int_{\nu_{LL}}^{\infty}\frac{L_\nu}{h\nu}d\nu
\label{eqU} 
\end{equation}
is the number of hydrogen ionizing photons per unit time, $\nu_{LL}$ is the frequency 
corresponding to the Lyman edge and r is the distance from the central source.
In the top panel of Fig.~\ref{modfig} we plot different column density ratios as a 
function of U.  The vertical short dashed line in the right hand side of the figure marks the 
upper limit of U (i.e. log~U$\le -$1.18).  Here we use the fact that N(H~{\sc i})/N(H) cannot be 
less than 10$^{-4}$.  The constraint on the total hydrogen column density, N(H), comes from the 
Chandra observations of Worrall et al. (2004).  They find no evidence for any intrinsic 
absorption in excess of 5$\times10^{19}$ cm$^{-2}$ in the direction of 3C48.  
%
A conservative lower limit of log~U=$-$1.82 is obtained from the column density ratios 
N(N~{\sc iv})/N(N~{\sc iii}) and N(S~{\sc vi})/N(S~{\sc iv}).  From Fig.~\ref{modfig}, 
it is apparent that the ratio N(C~{\sc iv})/N(N~{\sc iv}) is independent of U when log~U$\le-$1.0.  
As the gas is optically thin our choice of N(H~{\sc i}) and Solar abundance ratios, 
Z$_\odot$ will not alter these constraints.  
The observed ratio of N(C~{\sc iv})/N(N~{\sc iv}) is $\sim$0.1 dex below the predictions of our model.  
This is within 
the measurement uncertainties of our data.  Therefore, within the measurement 
uncertainties, [N/C] is very close to Solar values.    
This is in contrast with the systematically higher supersolar [N/C] ratios observed in 
most of the QSOs, especially the ones at high redshift or having high luminosities 
(Hamann \& Ferland 1992, 1993, 1999; Petitjean \& Srianand 1999).  
The rapid star-formation models enhancing the nitrogen abundance by the secondary CNO 
nucleosynthesis have been used to explain the [N/C]$>$[N/C]$_\odot$ ratios in these 
high redshift/luminosity systems.  
Using the observed flux density, 
F(1246.7\AA)=3.13$\times10^{-15}$ erg s$^{-1}$ cm$^{-2}$ \AA$^{-1}$ from the 
IUE data, a luminosity distance of 1950.7 Mpc and  a spectral index of 
$\alpha$=0.5, we estimate the luminosity, $\nu$L$_\nu$, at 1450 \AA~ 
for 3C48 to be 3.5$\times$10$^{11}$~L$_\odot$.
From Fig.~7 of Hamann et al. (1999), it can be seen that Solar values of 
[N/C] are not unusual for 3C48 with \zem=0.3700 and above estimated luminosity at 1450 \AA. 
%
In the bottom panel of Fig.~\ref{modfig}, we plot the predicted values of column densities 
of various species as a function of U.  Horizontal lines mark the observed values.  Clearly, 
observed column densities of \civ and \svi are lower by $\sim$0.5 dex compared to the model predictions 
within the allowed range of ionization parameter.  As the absorbing gas is 
optically thin, this difference can be accounted for by lowering
either the value of N(H~{\sc i}) or the metallicity.  The  
allowed range of N(H~{\sc i}) listed in Table~\ref{tcol}, constrains 
the metallicity of the absorbing gas to be in range  
0.1$\le$Z/Z$_\odot$$\le$1.3. 

The column density of  \neviii as predicted by the photoionization models is 
more than two orders of magnitude lower than the \neviii column density estimated
from the possible detection of this line. 
Reconciliation of the observed \neviii value requires neon abundance enhancement or 
opstulating regions with different levels of ionisation (multiply-ionized regions). 
Since the rest of the observed species can be explained by Solar 
abundance ratios it seems unlikely that only Ne has an unusual abundance. 
Thus the possible detection of \neviii in the poor signal-to-noise ratio FUSE data can be 
understood by the existence of multiply-ionized regions.  
Such multiply-ionized regions have been used to explain the observed column 
densities of highly ionised species (including \neviii) in J2233$-$606, UM675 and 3C288.1
(Petitjean \& Srianand 1999; Hamann et al. 1997; Hamann et al. 2000).  
Thus it is possible that the low-ionization species originate from the high-density (low-ionization) 
phase and high-ionization species \neviii originate from the low-density (high-ionization) phase 
of the absorption-line system.  
Our data does show some evidence that the absorption-line system consists of a number of 
components.  However, due to low spectral resolution and poor signal-to-noise ratio it is 
difficult to say if  \neviii arises from the same component as the other low-ionization species.  
The density of the 
absorbing gas or its distance from the quasar nucleus can, in principle, be constrained using 
equation~(\ref{eqU}) if either of these two quantities
and $Q(H^o)$ are known.  We estimate the luminosity at the Lyman limit, 
${\rm L_{LL}}=5.4\times10^{29}$ erg s$^{-1}$ Hz$^{-1}$ in the quasar rest frame 
using the observed flux density, 
F(1246.7\AA)=3.13$\times10^{-15}$ erg s$^{-1}$ cm$^{-2}$ \AA$^{-1}$ from the 
IUE data, a luminosity distance of 1950.7 Mpc and  a spectral index of 
$\alpha$=0.5. This translates into Q(H$^o$)=1.3$\times10^{56}$ photons s$^{-1}$.  
Substituting numerical values in equation~(\ref{eqU}) then gives
\begin{equation}
{\rm U} = \frac{3.4\times10^{44}}{{\rm nr}^2}.
\label{eqUn} 
\end{equation} 
Taking a nominal value of log~U=$-$1.5 from the photoionization models, 
nr$^2_{pc}$ = 1.2$\times10^{9}$ cm$^{-3}$ pc$^2$, where r$_{pc}$ is 
the distance from the central source in parsecs. 
However, with the available information, it is not possible to get 
useful estimates of these parameters. Assuming that the absorbing
cloud was pushed outwards by the radio jet $\gapp$10$^5$ yr ago (cf. Section 5.2), 
it would have travelled $\gapp$100 pc with a velocity of $\sim$ 1000 km s$^{-1}$.
This implies that the number density n$\lapp$10$^5$ cm$^{-3}$.

%
\subsection{Jet-cloud interaction?}
The complex radio structure of 3C48 has led to suggestions that
the jet in 3C48 may be disrupted by strong interaction with the 
external medium (e.g. Wilkinson et al. 1991; Feng et al. 2005).
This is not surprising as CSSs tend to be more asymmetric in both the 
separation ratio (r$_D$), defined to be the ratio of the separation from the 
core/nucleus of the farther component to the nearer one, and 
the flux density ratio (r$_L$), defined to be the ratio of integrated 
flux densities of the farther to the nearer component, 
compared with the larger sources, suggesting interaction 
with an asymmetric distribution of gas in the central regions 
(Saikia et al. 1995, 2001).  
The component nearer to the nucleus often tends to be the brighter one,
which is possibly a signature of the expansion of nearer component 
through the denser environment leading to higher dissipation of energy.
Orientation effects can also lead to substantial enhancements in these 
observed asymmetry parameters (cf. Jeyakumar et al. 2005 and references therein). 
Using the images and core position as given by Feng et al. (2005), who 
have recently reported the detection of a counter-jet in 3C48,  
the separation ratio for the prominent northern peak N2 and the counter-jet S 
is $\approx$2.8.  This could increase to $\approx$5 if the 
region of enhanced intensity in the outer region is taken instead of N2. 
The flux density ratio (r$_L$) of the northern or approaching 
components to the southern one, S, lies between $\sim$200 and 400
for observations at different frequencies and with different resolutions
(Feng et al. 2005). 
The source also appears significantly misaligned, possibly due to
collision with clouds of gas and also exhibits jitters in the direction of the
ejection axis as seen in the VLBA 8-GHz image of Worrall et al. (2004).
The misalignment angle, defined to be the supplement of the angle formed
at the core by N2 and S is about 30$^\circ$. For an inclination angle of 
30$-$35$^\circ$ for the jet, estimated from the relative strength of the
core, and an intrinsic misalignment of $\sim$10$-$15$^\circ$, r$_D$ is 
in the range of $\sim$3.8 to 5 for a velocity of advancement of $\sim$0.5c.
The corresponding flux density ratio, r$_L$=r$_D^{2+\alpha}$ (e.g. Scheuer \&
Readhead 1979; Blandford \& K\"{o}nigl 1979) is in the range of $\sim$50$-$100 
for a spectral index of 0.9,
which is lower than the observed flux density ratio by at least a factor of $\sim$3.
This difference could be attributed to intrinsic asymmetries in the gas 
distribution (cf. McCarthy, van Breugel \& Kapahi 1991). Using the formalism of 
Eilek \& Shore (1989) and Gopal-Krishna \& Wiita (1991), we estimate the density
ratio on opposite sides to be $\sim$7. However, these numbers should be treated with
caution because of the complex structure of the source and uncertainties in some
of the basic parameters such as the velocity of advancement.

The detection of blue-shifted absorption lines caused by clouds of
gas in the path of the radio jet would be an indication of
jet-cloud interaction.  Chatzichristou et al. (1999) report the detection 
of blue-shifted emission line ([O{\sc iii}]~$\lambda\lambda$4959,5007 and 
[O{\sc ii}]~$\lambda$3727) gas in 3C48 over 1.9 arcsec$^2$ of the 
central region.  On the basis of the velocity structure 
of the blue-shifted component of the emission line gas and its increasing strength in 
the same direction as the radio jet, they suggest that this gas contains the 
imprint of interaction with the jet.
The kinematical properties of this blue-shifted emission line gas, which shows clumpy 
morphology and 
has an outflow velocity of $\sim$580 \kms, are similar to the one detected in absorption.  
Using the mass estimate of $\sim10^9$M$_\odot$ (Fabian et al. 1987) for the emission line gas, 
determined from the strength of [O{\sc iii}]~$\lambda$5007 in 3C48, Wilkinson et al. (1991) 
show that energetics of the radio jet are capable of driving this gas with 
velocities as high as $\approx$1000 \kms. 
This is similar to the outflowing velocity of $\sim$1000 \kms for the blue-shifted material
seen in absorption. Although blue-shifted absorption can be caused by other physical
processes like radiative acceleration, the kinematical similarities with the emission-line
gas and the situation in 3C48 make jet-cloud interaction a likely possibility.

The interaction with the jet can impart momentum to the gas and can also heat as well 
as ionize it.  The gas ploughed and compressed by the jet can heat up to postshock 
temperatures of $\sim$10$^8$K (de Vries et al. 1999).  Taking adiabatic and radiative 
cooling into account de Vries et al. find that such a gas with a density of 
1$-$5 cm$^{-3}$ will cool to 
10$^4$K in about 30,000 yr. Using the recombination coefficients from P\'{e}quignot, 
Petitjean \& Boisson  (1991), we find that the recombination timescales for gas with 
a temperature few times 10$^4$K and density 5 cm$^{-3}$ are only a few thousand years 
for most of the species considered here.  The maximum recombination timescale occurs 
for \hi and O~{\sc i}, which we estimate to be $\sim$10$^4$ yr.  
Therefore, any thermal or ionization signature of jet$-$cloud interaction will be lost
in timescales $\gapp$10$^5$ yr.  Since we can explain the observed column densities 
of the absorbing cloud using the photoionization models alone, 
any injection of energy resulting from a jet-cloud interaction must have taken place 
before 10$^5$ yr.

\section {Summary}
We have reported the detection of a \zabs=0.3654 absorption-line system 
in the UV spectrum of the CSS quasar 3C48, which has a highly distorted radio
structure.  The absorbing material is blueshifted by $\sim$1000 \kms with respect 
to \zem=0.3700 estimated from the optical emission lines. The outflowing material shows 
absorption lines from Ly$\beta$, Ly$\gamma$, C~{\sc iv}, S~{\sc vi}, O~{\sc vi}, 
N~{\sc iv} and possibly O~{\sc iv} and Ne~{\sc viii}.  
Detection of high-ionization lines such as \ovi and \neviii  suggests that the 
outflowing material is located close to the nucleus of the quasar.  
The kinematical properties of the \zabs=0.3654 system are very similar 
to the blue-shifted emission line gas detected by Chatzichristou et al. (1999).
The complex radio structure has been interpreted to be due to strong interaction 
with the ambient medium.  On the 
basis of spatial correlation between the radio source and emission line gas, Chatzichristou 
et al. argue that this component bears the imprint of the interaction between the jet and the  
ambient gas. Considering the similarity of the kinematics of the emitting and absorbing material 
and the physical situation in 3C48, it is  possible that the outflowing 
material detected in absorption may have also interacted with the radio jet. 

Interaction of a cloud with a radio jet can  impart momentum and as well as heat and 
ionize it.  The recombination and cooling timescales suggest that signatures of such a
jet-cloud interaction will be lost on time scales $\gapp$10$^5$ yr.  
We find that photoionization models with Solar abundance 
ratios (with overall metallicity in the range 0.1$\le$Z/Z$_\odot$$\le$1.3) 
are enough to explain the 
observed column densities of all the species except for \neviii.
The consistency with the photoionization-models suggests that the 
any possible interaction of absorbing material with the jet must have 
taken place $\gapp$10$^5$ yr ago.
Reconciliation of the possible \neviii detection requires 
regions with different levels of ionization. We also find that
the abundance ratio of nitrogen to carbon is close to Solar values,
unlike in the case of most quasars, especially at high-redshifts, 
which have super-Solar values. 

Our GMRT spectra of 3C48, from which we estimate the 3$\sigma$ upper 
limit to the optical depth to lie in the range 0.001 to 0.003, is consistent with 
the low N(H~{\sc i}) detected towards the optical point source.  
Assuming a spin temperature of 100 K and line width of 100 \kms, these 
correspond to column densities of
 N(H~{\sc i})$\le$(1.8$-$5.4$)\times10^{19}$ cm$^{-2}$.  Due to the diffuse nature of 
the radio structure of 3C48, \hi column densities as high as 10$^{21}$ cm$^{-2}$ towards 
individual knots or compact features cannot be ruled out.

%
%
\section*{Acknowledgments}
We thank an anonymous referee for several useful comments and suggestions.
We are grateful to MAST archive for the excellent user support and documentation.  We 
also thank the numerous contributers to GNU/Linux group. This research has made use of 
the NASA/IPAC Extragalactic Database (NED) which is operated by the Jet Propulsion Laboratory, 
California Institute of Technology, under contract with the National Aeronautics and Space 
Administration.  
One of us (NG) thanks the Kanwal Reiki Scholarship of TIFR for partial financial support.   
We also thank the GMRT staff for their cooperation during our GMRT observations.  The GMRT 
is a national facility operated by the National Centre for Radio Astrophysics of the 
Tata Institute of Fundamental Research.  
%

\label{lastpage}

\end{document}